\documentclass[twocolumn,showpacs,preprintnumbers,amsmath,amssymb]{revtex4}

\usepackage{graphicx}
\usepackage{dcolumn}
\usepackage{bm}

\newcommand{\bfb}{\mbox{\boldmath$b$}}

\newcommand{\bfk}{\mbox{\boldmath$k$}}
\newcommand{\bfm}{\mbox{\boldmath$m$}}
\newcommand{\bfu}{\mbox{\boldmath$u$}}
\newcommand{\bfx}{\mbox{\boldmath$x$}}
\newcommand{\bfzhat}{\mbox{\boldmath $ {\hat z} $ } }
\newcommand{\bfB}{\mbox{\boldmath$B$}}

\newcommand{\bfG}{\mbox{\boldmath$G$}}

\newcommand{\bfU}{\mbox{\boldmath$U$}}

\newcommand\calD{{\cal D}}
\newcommand\calE{{\cal E}}
\newcommand\calL{{\cal L}}
\newcommand{\bfcalE}{\boldsymbol{{\cal E}}}
\newcommand{\bfcalEhat}{\hat{\boldsymbol{{\cal E}}}}

\newcommand{\bfkhat}{\mbox{\boldmath $\hat k$}}
\newcommand{\bfxhat}{\mbox{\boldmath $\hat x$}}
\newcommand{\bfbhat}{\mbox{\boldmath $\hat b$}}

\newcommand{\bfBhat}{\mbox{\boldmath $\hat B$}}
\newcommand{\bfGhat}{\mbox{\boldmath $\hat G$}}

\begin{document}

\preprint{APS/123-QED}

\title{The Turbulent Magnetic Diffusivity Tensor for Time-Dependent Mean Fields}

\author{David W. Hughes}
\altaffiliation[Permanent address: ]{Department of Applied Mathematics, University of Leeds, Leeds LS2 9JT, UK}
\email{d.w.hughes@leeds.ac.uk}
\author{Michael R.E. Proctor}
\altaffiliation[Permanent address: ]{Centre for Mathematical Sciences, University of Cambridge, Wilberforce Road, Cambridge CB3 0WA, UK}
\email{mrep@cam.ac.uk}
\affiliation{Institut Henri Poincar\'e, 11 rue Pierre et Marie Curie, 75005 Paris, France}

\date{\today}

\begin{abstract}
We re-examine the nature of the turbulent magnetic diffusivity tensor of mean field electrodynamics and show that an inconsistency arises if it is calculated via consideration of time-independent magnetic fields. Specifically, the predicted growth rate of the mean magnetic field is, in general, incorrect. We describe how the traditional expansion procedure for the mean electromotive force should be extended, and illustrate the consistency of this approach by means of a perturbation analysis for a mean magnetic field varying on long spatial scales. Finally we examine the magnitude of this new contribution to the magnetic diffusion for a particular flow.
\end{abstract}

\maketitle







Large-scale magnetic fields are observed in a vast range of cosmical bodies, from planets to stars and accretion discs. Understanding their generation by turbulent dynamo action remains one of the most challenging problems in astrophysics. Cosmical dynamos are often studied within the framework of mean field electrodynamics, a tremendously elegant theory that describes the evolution of a mean (large-scale) magnetic field in terms of transport coefficients determined from averaged small-scale properties of the flow and field. The main underlying assumption of mean field electrodynamics is one of scale separation for the velocity and magnetic fields, which is nearly always understood as \textit{spatial} scale separation. In this case, the velocity and magnetic field can formally be decomposed into mean and fluctuating parts,
\begin{equation}
\bfU = \bfU_0 + \bfu , \qquad
\bfB = \bfB_0 + \bfb ,
\label{eq:meanfluc}
\end{equation}
where the fluctuating quantities $\bfu$ and $\bfb$ vary on the (small) scale $\ell$, which may be regarded as a typical turbulent eddy size; the mean fields $\bfU_0$ and $\bfB_0$ vary on some much larger scale $L \gg \ell$. Mean field electrodynamics is, at heart, a kinematic theory; here the velocity $\bfU ( \bfx , t )$ is assumed to be known, with the magnetic field determined solely by the induction equation
\begin{equation}
\frac{\partial \bfB}{\partial t} = \nabla \times \left( \bfU \times \bfB \right) + \eta \nabla^2 \bfB ,
\label{eq:induction}
\end{equation}
where $\eta$ is the magnetic diffusivity (here assumed uniform). In this paper, in which we shall address a fundamental issue in the interpretation of mean field electrodynamics, we shall also adopt this kinematic approach. We shall not consider the dynamical back-reaction of the magnetic field on the flow via the Lorentz force, important though it is in natural dynamos.

Averaging equation~(\ref{eq:induction}) over some intermediate scale $a$, satisfying $\ell \ll a \ll L$, leads to the following equations for the mean and fluctuating magnetic fields,
\begin{equation}
\frac{\partial \bfB_0}{\partial t} = \nabla \times \left( \bfU_0 \times \bfB_0 \right) 
+ \nabla \times \bfcalE
+ \eta \nabla^2 \bfB_0 ,
\label{eq:mean_induction}
\end{equation}
\begin{equation}
\frac{\partial \bfb}{\partial t} = \nabla \times \left( \bfU_0 \times \bfb \right) 
+ \nabla \times \left( \bfu \times \bfB_0 \right)
+ \nabla \times \bfG
+ \eta \nabla^2 \bfb ;
\label{eq:fluc_induction}
\end{equation}
here $\langle \ \, \rangle$ denotes a spatial average, $\bfcalE = \langle \bfu \times \bfb \rangle$ is the mean electromotive force (emf) and $\bfG = ( \bfu \times \bfb ) - \langle \bfu \times \bfb \rangle $.

Equation~(\ref{eq:fluc_induction}) can be written formally as
\begin{equation}
\calL ( \bfb ) = \nabla \times \left( \bfu \times \bfB_0 \right) ,
\label{eq:fluc_op}
\end{equation}
where $\calL$ is a linear operator. The traditional approach to tackling (\ref{eq:fluc_op}) \cite{Moff,KR80} assumes that if there is no mean field $\bfB_0$ then there are no non-decaying solutions for $\bfb$; in other words, all solutions to the homogeneous equation $\calL (\bfb) = 0$ decay. In reality, at high magnetic Reynolds number $Rm$ this is unlikely to be the case; in other words, \textit{small-scale} dynamo action will ensue. This can lead to severe difficulties in the consistency of the mean field approach, as discussed in \cite{ch09}. However, in this Letter, we shall not explore this particular issue. Instead, we shall assume, as is certainly the case for low $Rm$, that the fluctuating magnetic field $\bfb$ is indeed driven entirely by a non-zero right hand side of equation~(\ref{eq:fluc_op}). It then follows that $\bfb$, and hence $\bfcalE$, is linearly and homogeneously related to the mean field $\bfB_0$. Accordingly, one may posit an expansion for $\bfcalE$ in terms of $\bfB_0$ and its derivatives. This is always written as
\begin{equation}
{\calE}_i = \alpha_{ij}B_{0j} + \beta_{ijk}\frac{\partial B_{0j}}{\partial x_k} + \ldots ,
\label{eq:mean_emf}
\end{equation}
where it is anticipated that the large spatial scale of $\bfB_0$ will lead to rapid convergence.

The aim of mean field electrodynamics is first to calculate the coefficients of the tensors $\alpha_{ij}$ and $\beta_{ijk}$, and then to substitute for $\bfcalE$ into (\ref{eq:mean_induction}) to obtain an equation determining the temporal evolution of a large-scale field. It is the inconsistency between these two steps that we shall explore in this Letter. In general, expression~(\ref{eq:mean_emf}) should also contain temporal derivatives of $\bfB_0$, although, as pointed out in \cite{Moff}, these may be replaced by spatial derivatives by using back-substitution from (\ref{eq:mean_induction}). However, and this is the crux of our argument, although this manipulation is formally correct, it leads to an inconsistency in the determination of the evolution of the mean field when the expression for $\bfcalE$ is substituted from (\ref{eq:mean_emf}) into (\ref{eq:mean_induction}).

The difficulty arises as a consequence of the natural means of determining the tensors $\alpha_{ij}$ and $\beta_{ijk}$ in expression~(\ref{eq:mean_emf}). Given a turbulent flow $\bfu ( \bfx ,t)$, the components $\alpha_{ij}$ can be determined by evaluation of the mean emf $\bfcalE$ following the imposition of three independent, steady, spatially uniform fields $\bfB_0$. Having calculated all the components of the $\alpha_{ij}$ tensor, the components 
$\beta_{ijk}$ can subsequently be determined by consideration of the emf following the imposition of independent, steady fields having a uniform \textit{gradient}. Substitution into (\ref{eq:mean_induction}) then leads to an expression for the growth rate of the mean field in terms of the $\alpha$ and $\beta$ tensors. However, in general, \textit{this expression will not provide the correct description of the evolution of a large-scale field.} The problem arises because it is inconsistent to calculate $\beta_{ijk}$ from a time-independent spatially-dependent field, since, in reality, a field with spatial dependence will vary with time. To obtain a correct representation for the growth rate, premature back-substitution for the temporal derivatives of the mean field must be avoided. Instead of (\ref{eq:mean_emf}), the following expansion should be used for the mean emf:
\begin{equation}
{\calE}_i = \alpha_{ij}B_{0j} + \Gamma_{ij}\frac{\partial B_{0j}}{\partial t} + {\beta}_{ijk}\frac{\partial B_{0j}}{\partial x_k} + \ldots \, ,
\label{eq:mean_emf_2}
\end{equation}
where the coefficients $\alpha_{ij}$ and $\beta_{ijk}$ are identical to those in expression (\ref{eq:mean_emf}) and can be calculated precisely as described above, using time-independent magnetic fields. The new tensor $\Gamma_{ij}$ is to be determined via evaluation of $\bfcalE$ after imposition of a spatially uniform magnetic field that increases linearly with time. Substitution of the expression for $\bfcalE$ from (\ref{eq:mean_emf_2}) into (\ref{eq:mean_induction}) (ignoring both the mean flow and the molecular diffusion term, neither of which is important for the argument we are advancing here) yields an equation of the form
\begin{equation}
\frac{\partial B_{0i}}{\partial t} = \epsilon_{ijk} \frac{\partial}{\partial x_j}
\left( \alpha_{km} B_{0m} + \Gamma_{km} \frac{\partial B_{0m}}{\partial t} 
+ \beta_{kmn}\frac{\partial B_{0m}}{\partial x_n} \right) .
\label{eq:mean_induction_2}
\end{equation}
Making use of the fact that the expression for the emf is assumed to be a rapidly convergent series, we may \textit{now} back-substitute for the time derivative of $B_0$ on the right hand side of (\ref{eq:mean_induction_2}) using just the leading order terms, thus yielding
\begin{eqnarray}
\nonumber
\frac{\partial B_{0i}}{\partial t} &=& \epsilon_{ijk} \frac{\partial}{\partial x_j}
\left( \alpha_{km} B_{0m}  \qquad \qquad \right. \\
 &+& \Gamma_{km} \epsilon_{mpq} \frac{\partial}{\partial x_p} (\alpha_{qr}B_{0r}) 
+ \beta_{kmn}\frac{\partial B_{0m}}{\partial x_n} \left . \right) .
\label{eq:mean_induction_3}
\end{eqnarray}
The new term involving $\Gamma_{km}$ has one spatial derivative and hence is of the same order as that involving $\beta_{kmn}$. For simplicity, if we consider the case when the components of the $\alpha$, $\beta$ and $\Gamma$ tensors are constants (more generally, they could be functions of the slow spatial variation), it can be seen that the coefficient of turbulent diffusivity (more precisely, the coefficient of the second order spatial derivative term) is not $\beta_{ijk}$ but is instead
\begin{equation}
\epsilon_{mkq} \alpha_{qj} \Gamma_{im} + \beta_{ijk} .
\label{eq:diff_correct}
\end{equation}

As a complementary approach, which reinforces the above arguments, we consider, via a classical perturbation analysis, the evolution, under a small-scale velocity field, of a magnetic field with a long-wavelength modulation; this is the approach adopted by Roberts \citep{Rob70} in his analysis of two-dimensional cellular flows. We write the mean field in the form $\bfB_0 = \bfBhat_0 \exp(i \bfk \cdot \bfx + pt)$ and the fluctuating field as $\bfb = \bfbhat \exp(i \bfk \cdot \bfx + pt)$, where $\bfbhat$ varies on the same small spatial and temporal scales as $\bfu$. The wavenumber $k = | \bfk |$ is assumed to be small; accordingly, we may develop the solution in powers of $k$; specifically, the growth rate has the expansion 
\begin{equation}
p=k p_1 + k^2 p_2 + \cdots .
\label{eq:p}
\end{equation}
The coefficient $p_2$ gives an unambiguous description of the diffusion term.

At leading order, the fluctuating and mean induction equations take the form
\begin{eqnarray}
\label{eq:fluc_k0}
(\partial_t - \eta \nabla^2) \bfbhat_0 &=& \bfBhat_0 \cdot \nabla \bfu + \nabla \times \bfGhat_0 , \\
p_1 \bfBhat_0 &=& i \bfkhat \times \bfcalEhat_0 ;
\label{eq:mean_k0}
\end{eqnarray}
and at the next order,
\begin{eqnarray}
\nonumber (\partial_t - \eta \nabla^2) \bfbhat_1 + p_1 \bfbhat_0 - 2 i \eta \bfkhat \cdot \nabla \bfbhat_0 &=&\\
\label{eq:fluc_k1}  - i (\bfu \cdot {\bfkhat}) \bfBhat_0 + \nabla \times \bfGhat_1 &+& i \bfkhat \times \bfGhat_0 , \\
\label{eq:mean_k1}
p_2 \bfBhat_0 = i \bfkhat \times \bfcalEhat_1 &-& \eta \bfBhat_0
\end{eqnarray}
(where we have assumed $\nabla \cdot \bfu = 0$ for simplicity). The formal means of solution is now clear, although the inversion of the various linear operators may of course lead to technical difficulties. The fluctuating field $\bfbhat_0$ can be obtained from (\ref{eq:fluc_k0}); $\bfcalEhat_0$ can then be evaluated and hence, from (\ref{eq:mean_k0}), $p_1$ determined as
\begin{equation}
p_1 = \frac{\left( i \bfkhat \times \bfcalEhat_0 \right) \cdot \bfBhat_0^*}{B_0^2} .
\label{eq:p1}
\end{equation}
Proceeding, equation~(\ref{eq:fluc_k1}) can be solved (formally) for $\bfbhat_1$; $\bfcalEhat_1$ can then be evaluated and hence $p_2$ calculated, from (\ref{eq:mean_k1}), as
\begin{equation}
p_2 = \frac{\left( i \bfkhat \times \bfcalEhat_1 - \eta \bfBhat_0 \right) \cdot \bfBhat_0^*}{B_0^2} .
\label{eq:p2}
\end{equation}
The traditional diffusive term (i.e.\ that represented by the $\beta$ tensor) comes not from solving (\ref{eq:fluc_k1}), but instead from solving the equation
\begin{equation}
(\partial_t - \eta \nabla^2) \bfbhat_1 = - i (\bfu \cdot {\bfkhat}) \bfBhat_0 + \nabla \times \bfGhat_1 .
\label{eq:fluc_k1_wrong}
\end{equation}
It is clear that this does not tell the whole story regarding the growth rate; the term $p_1 \bfbhat_0$ in (\ref{eq:fluc_k1}), the effects of which are encapsulated in the $\Gamma$ tensor, may be equally important.

We may be more explicit if we consider a simple example in which $\bfu$ (and hence $\bfbhat$) takes the form
$\bfu = \mathbb{R} ({\tilde \bfu} \exp (i(\bfm \cdot \bfx - \omega t))$. The awkward $\bfG$ terms then vanish, allowing explicit solution of equations (\ref{eq:fluc_k0}) and (\ref{eq:fluc_k1}) for all $Rm$.
Then $\bfcalEhat_0$ is given by
\begin{equation}
\bfcalEhat_0  = \frac{1}{2} \, \mathbb{R} \left\{ \frac{{\tilde \bfu^*} \times (i \bfm \cdot \bfBhat_0 ) {\tilde \bfu}}{\calD} \right\} ,
\label{eq:E0}
\end{equation}
where $\calD = -i \omega + \eta m^2$. The quantity $p_1$ then follows from expression (\ref{eq:p1}). At the next order we obtain
\begin{eqnarray}
\nonumber 
\bfcalEhat_1  &=&  - \frac{1}{2} \, \mathbb{R} \left\{ 
\frac{i {\tilde \bfu^*} \times ({\tilde \bfu} \cdot \bfkhat) \bfBhat_0}{\calD} \right. \\
\nonumber
&+& \frac{2 \eta (\bfm \cdot \bfkhat) {\tilde \bfu^*} \times (i \bfm \cdot \bfBhat_0 ) {\tilde \bfu} }{\calD^2} \\
&+& \left. \frac{p_1 {\tilde \bfu^*} \times (i \bfm \cdot \bfBhat_0 ) {\tilde \bfu} }{\calD^3}
\right\} ,
\label{eq:E1}
\end{eqnarray}
with $p_2$ given by expression (\ref{eq:p2}). The first two terms are encapsulated in the traditional $\beta_{ijk}$ term. However, the final term, being proportional to $p_1$, can be identified with the first term in (\ref{eq:diff_correct}). This term arises from the fact that we are solving an eigenvalue problem for the growth rate, rather than studying the statistically steady response to an imposed current. We note here that the final term in expression (\ref{eq:E1}) is small compared with the other terms when $Rm$ is small. If for example we regard $\eta$ as fixed and thus take $Rm\propto |\bfu|$, we see that the last term $\propto p_1 Rm^2\sim Rm^4$, while the other terms $\propto Rm^2$ when $Rm$ is small. In this situation the First Order Smoothing Approximation can be used for quite general velocity fields, leading to formulae analogous to (\ref{eq:E0},\ref{eq:E1}). However when $Rm$ is not small, the new term may be at least as large as the others, while direct calculation is not possible.

It is therefore of interest to investigate, for a specific flow, the relative magnitudes of the two contributions to the diffusivity in expression (\ref{eq:diff_correct}), particularly when $Rm$ is much greater than unity. This can be most readily achieved by considering spatially-periodic velocities of the form $\bfu (x,y,t)$; for such flows, the magnetic field takes the form $\bfB ( \bfx ,t) = \bfBhat (x,y) \exp(pt+ikz)$. By consideration of the kinematic dynamo problem for a range of small values of $k$, one can obtain the first two terms in the expansion for the growth rate,
\begin{equation}
p = p_1 k + p_2 k^2 .
\label{eq:pk}
\end{equation}
The real part of $p_1$ provides information on $\alpha_{ij}$, the real part of $p_2$ encompasses all of the diffusive contributions. Via the complementary approach of imposing two independent spatially uniform magnetic fields of the form, for example, 
\begin{equation}
\bfB_0 = (B_0 + C_0t,0,0) \ \textrm{and} \ 
\bfB_0 = (0, B_0 + C_0t,0),
\label{eq:impfield}
\end{equation}
constraining the perturbation magnetic field to be independent of $z$ (thus eliminating the possibility of small-scale dynamo action), and then taking spatial averages over the $xy$-plane and temporal averages over the fast flow time scale, we may independently calculate the tensors $\alpha_{ij}$ and $\Gamma_{ij}$ (here $1 \le i,j \le 2$). We are then in a position to quantify the two contributions to the magnetic diffusivity in expression (\ref{eq:diff_correct}). We should note that, at least in theory, it is also possible to obtain an independent measure of the elements of the $\beta_{ijk}$ tensor, by calculating the emf after the imposition of steady magnetic fields with a uniform gradient. However, this is extremely computationally expensive, in comparison with the calculations of the $\alpha$ and $\Gamma$ terms, since it requires a spatial domain that is very large compared with a typical velocity scale; by contrast, the computational domain needed to calculate the $\alpha$ and $\Gamma$ terms is simply that of the spatial periodicity of the flow.

Here we consider the specific two-dimensional time-dependent flow introduced by Otani \citep{Otani_93} (sometimes referred to as the MW+ flow), given by
\begin{equation}
\bfu(x,y,t) = \nabla \times ( \psi(x,y,t) \bfzhat) + \psi(x,y,t) \bfzhat,
\label{eq:2dflow}
\end{equation}
where
\begin{equation}
\psi(x,y,t)= 2 \cos^2 t \cos x - 2 \sin^2 t \cos y .
\label{eq:otani}
\end{equation}
Symmetry considerations \cite{CG95,Cour_08} dictate that $\alpha_{ij}$ and $\beta_{ijk}$ take the form $\alpha_{ij}=\alpha \delta_{ij}$, $\beta_{ijk}=\beta \epsilon{ijk}$; similarly it can be shown that $\Gamma_{ij}= \Gamma \delta_{ij}$. Thus we need only consider an imposed field in either the $x$ or $y$-direction in order to determine $\alpha_{ij}$ and $\Gamma_{ij}$. It follows from (\ref{eq:diff_correct}) that the `new' contribution to the diffusivity is $- \alpha \Gamma$.

\begin{figure}
\includegraphics[scale=0.65]{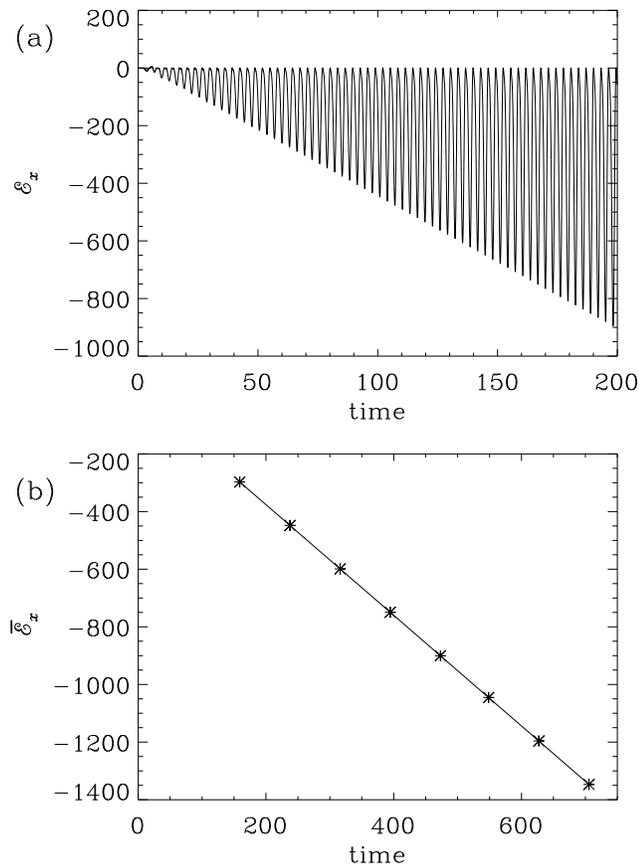}
\caption{\label{fig} (a)~Spatially averaged emf, $\calE_x$, versus time for the MW+ flow, for $Rm=100$ and an imposed field $\bfB_0 = t \bfxhat$. (b)~$\bar \calE_x$, obtained from averaging $\calE_x$ over the fast time scale of the flow, versus time.}
\end{figure}

Figure~\ref{fig}(a) shows the spatially, but not temporally, averaged emf $\calE_x$ (i.e.\ $\langle \bfu \times \bfb \rangle_x$) versus time, for an imposed field $\bfB_0 = t \bfxhat$ and for $Rm=100$. Figure~\ref{fig}(b) shows $\bar \calE_x$, the emf after a further temporal average over the fast time scale of the flow. Since, for an imposed field of the form $\bfB_0 = (B_0+C_0t) \bfxhat$, $\bar \calE_x$ takes the form
\begin{equation}
{\bar \calE_x} = \alpha (B_0 +C_0 t) + \Gamma C_0 ,
\label{eq:emfx}
\end{equation}
$\alpha$ and $\Gamma$ can be evaluated from the slope and intercept of the straight line in Fig.\ \ref{fig}(b). (Note that we can obtain both quantities simply by having $C_0$ non-zero; setting $B_0 \ne 0$, $C_0=0$ (the traditional means of computing $\alpha$) supplies an independent verification of $\alpha$).

Table~\ref{tab:table1} shows, for a range of $Rm$, the values of $p_1$ and $p_2$, obtained from solving the kinematic dynamo problem for a range of small values of $k$, together with the values of $\alpha$ and $\Gamma$, determined from the emf after the imposition of a spatially uniform field $\bfB_0 = t \bfxhat$. The final column contains the $-\alpha \Gamma$ contribution to the diffusivity; the full diffusivity is given by $-p_2 = \beta - \alpha \Gamma$. (By the very nature of the means of determining these values, with the $p_j$ obtained from fitting a parabola to the $p$ versus $k$ curve, whereas $\alpha$ and $\Gamma$ are determined from long temporal averages, the former (particularly $p_2$) cannot be determined to the same level of accuracy as the latter.)

An interesting pattern emerges from Table~\ref{tab:table1}. At low values of $Rm$ the diffusivity is predominantly due to $\beta$; this is consistent with the explicit result for the simple monochromatic flow discussed above. However, as $Rm$ is increased, the $-\alpha \Gamma$ term increases in importance until, at $Rm=100$ it is the dominant contribution. For $Rm =50$ and $100$ the value of $\beta$ is negative, though small in magnitude in comparison with the total diffusivity $-p_2$. This is certainly possible, though it should also be borne in mind that the determination of $p_2$ involves some slight inaccuracy. The crucial element though, which is unaffected by small changes in $p_2$, is that the magnetic diffusion at these higher values of $Rm$ comes almost completely from the $-\alpha \Gamma$ contribution.

\begin{table}
\caption{\label{tab:table1} $p_1$ and $p_2$, calculated from the $p$ vs $k$ curve after solving the kinematic dynamo problem, together with $\alpha$ and $\Gamma$, calculated from consideration of the emf after the imposition of the mean field $\bfB_0 = t \bfxhat$, for a range of values of $Rm$. }
\begin{ruledtabular}
\begin{tabular}{cccccc}
$Rm$ & $p_1$ & $p_2$ & $\alpha$ & $\Gamma$ & - $\alpha \Gamma$  \\
\hline
1    &  0.80  & -1.56  &  -0.80  &  0.20  & 0.16 \\
10   &  0.75  & -0.93  &  -0.75  &  0.85  & 0.64 \\
50   & 1.51   & -6.2   &  -1.52  &  4.4   & 6.69 \\
100  & 1.9    & -14.5  &  -1.92  &  7.8   & 14.98 \\
\end{tabular}
\end{ruledtabular}
\end{table}

To summarize, we have identified an inconsistency in the usual means of expressing the mean emf only in terms of spatial derivatives. In the general expression for the emf it is vital to retain temporal derivatives of the mean field; substitution in terms of spatial derivatives can be performed, but this needs to be done in a consistent order-by-order manner. At small values of $Rm$ the new term contribution to the magnetic diffusivity is small; however, in the example we have studied, it becomes dominant at large $Rm$.

For clarity we have focused in this paper on a simple mean field problem with no mean flow. We could straightforwardly have included this term in the equations, which would have changed the dynamo growth rate and hence the value of the turbulent diffusivity term ($p_2$). A consequence of the work described here is that even if the mean flow can be ignored in the fluctuating equations, it may still affect the diffusivity since this is linked to the growth rate.

\begin{acknowledgments}
This work was performed at the Institut Henri Poincar\'e; we thank Emmanuel Dormy, Stephan Fauve and Fran\c{c}ois Petr\'elis for inviting us to attend. We are also grateful to Alice Courvoisier for helpful discussions. This research was supported by STFC and by a Royal Society Leverhulme Trust Senior Research Fellowship (DWH).
\end{acknowledgments}

\bibliography{HP_PRL_09b}

\begin{thebibliography}{7}
\expandafter\ifx\csname natexlab\endcsname\relax\def\natexlab#1{#1}\fi
\expandafter\ifx\csname bibnamefont\endcsname\relax
  \def\bibnamefont#1{#1}\fi
\expandafter\ifx\csname bibfnamefont\endcsname\relax
  \def\bibfnamefont#1{#1}\fi
\expandafter\ifx\csname citenamefont\endcsname\relax
  \def\citenamefont#1{#1}\fi
\expandafter\ifx\csname url\endcsname\relax
  \def\url#1{\texttt{#1}}\fi
\expandafter\ifx\csname urlprefix\endcsname\relax\def\urlprefix{URL }\fi
\providecommand{\bibinfo}[2]{#2}
\providecommand{\eprint}[2][]{\url{#2}}

\bibitem[{\citenamefont{Moffatt}(1978)}]{Moff}
\bibinfo{author}{\bibfnamefont{H.~K.} \bibnamefont{Moffatt}},
  \emph{\bibinfo{title}{Magnetic Field Generation in Electrically Conducting
  Fluids}} (\bibinfo{publisher}{Cambridge University Press},
  \bibinfo{year}{1978}).

\bibitem[{\citenamefont{Krause and R{\"a}dler}(1980)}]{KR80}
\bibinfo{author}{\bibfnamefont{F.}~\bibnamefont{Krause}} \bibnamefont{and}
  \bibinfo{author}{\bibfnamefont{K.-H.} \bibnamefont{R{\"a}dler}},
  \emph{\bibinfo{title}{Mean-Field Magnetohydrodynamics and Dynamo Theory}}
  (\bibinfo{publisher}{Pergamon}, \bibinfo{year}{1980}).

\bibitem[{\citenamefont{Cattaneo and Hughes}(2009)}]{ch09}
\bibinfo{author}{\bibfnamefont{F.}~\bibnamefont{Cattaneo}} \bibnamefont{and}
  \bibinfo{author}{\bibfnamefont{D.~W.} \bibnamefont{Hughes}},
  \bibinfo{journal}{Mon. Mot. R. Astron. Soc.}  (\bibinfo{year}{2009}).

\bibitem[{\citenamefont{Roberts}(1970)}]{Rob70}
\bibinfo{author}{\bibfnamefont{G.~O.} \bibnamefont{Roberts}},
  \bibinfo{journal}{Phil. Trans. R. Soc. Lond.}
  \textbf{\bibinfo{volume}{A266}}, \bibinfo{pages}{535} (\bibinfo{year}{1970}).

\bibitem[{\citenamefont{Otani}(1993)}]{Otani_93}
\bibinfo{author}{\bibfnamefont{N.~F.} \bibnamefont{Otani}},
  \bibinfo{journal}{J. Fluid Mech.} \textbf{\bibinfo{volume}{253}},
  \bibinfo{pages}{327} (\bibinfo{year}{1993}).

\bibitem[{\citenamefont{Childress and Gilbert}(1995)}]{CG95}
\bibinfo{author}{\bibfnamefont{S.}~\bibnamefont{Childress}} \bibnamefont{and}
  \bibinfo{author}{\bibfnamefont{A.~D.} \bibnamefont{Gilbert}},
  \emph{\bibinfo{title}{Stretch, Twist, Fold: the Fast Dynamo}}
  (\bibinfo{publisher}{Springer Verlag}, \bibinfo{year}{1995}).

\bibitem[{\citenamefont{Courvoisier}(2008)}]{Cour_08}
\bibinfo{author}{\bibfnamefont{A.}~\bibnamefont{Courvoisier}},
  \bibinfo{journal}{Geophys. Astrophys. Fluid Dyn.}
  \textbf{\bibinfo{volume}{102}}, \bibinfo{pages}{217} (\bibinfo{year}{2008}).

\end{thebibliography}

\end{document}